\documentclass[structabstract]{aa}

\usepackage{epsfig}
\usepackage{graphicx}
\usepackage{subfigure}
\usepackage{latexsym}
\usepackage{amssymb}
\usepackage{amsmath}
\usepackage{longtable,multirow}
\usepackage{color}
\usepackage{txfonts}
\usepackage{appendix}
\usepackage{ulem}
\usepackage{footnote}
\usepackage{natbib,twoopt}
\usepackage{booktabs}

\bibstyle{aa}

\begin{document}
\sloppy

\title{{\it Herschel}\thanks{{\it Herschel} is an ESA space observatory with science instruments provided by European-led Principal Investigator consortia and with important participation from NASA.} spectral-mapping of the Helix Nebula (NGC 7293):}
\subtitle{Extended CO photodissociation and OH$^+$ emission}

\author{M. Etxaluze\inst{1,2}\thanks{email: m.etxaluze@icmm.csic.es}
  \and J. Cernicharo\inst{1,2}
  \and J.~R. Goicoechea\inst{1,2}
  \and P.~A.~M. van Hoof\inst{3}
  \and B.~M. Swinyard\inst{4,5}
\and M.~J. Barlow\inst{4}\and\\ 
 G.~C. van de Steene\inst{3}
 \and M.~A.~T. Groenewegen\inst{3}
\and F. Kerschbaum\inst{6}
\and T. L. Lim\inst{5}
\and F. Lique\inst{7}\and\\
M. Matsuura\inst{4}
\and C. Pearson\inst{5,8}
\and E.~T. Polehampton\inst{5,9}
\and P. Royer\inst{10}
\and T. Ueta\inst{11,12}
}
\institute{Departamento de Astrof\'isica. Centro de Astrobiolog\'ia. CSIC-INTA. Torrej\'on de Ardoz, 28850 Madrid, Spain.
\and Astrochemistry group, Instituto de Ciencias de Materiales de Madrid, Cantoblanco, 28049 Madrid, Spain
\and Royal Observatory of Belgium, Ringlaan 3, B-1180 Brussels, Belgium
\and Department of Physics \& Astronomy, University College London, London WC1E 6BT, UK 
\and RAL Space, Rutherford Appleton Laboratory, Oxfordshire, OX11 0QX, UK 
\and University of Vienna, Department of Astrophysics, T\"urkenschanzstra\ss{}e 17, 1180 Wien, Austria
\and LOMC-UMR 6294, CNRS-Universit\'e du Havre, 25 rue Philippe Lebon, BP 540 76058 Le Havre France
\and Department of Physics and Astronomy, The Open University, Milton Keynes, MK7 6AA, UK
\and Institute for Space Imaging Science, Dept. of Physics \& Astronomy, University of Lethbridge, Lethbridge, AB T1K3M4, Canada
\and Institute of Astrophysics, KU Leuven, Celestijnenlaan 200D, 3001 Leuven, Belgium
\and Department of Physics and Astronomy, University of Denver, 2112 E. Wesley Ave., Denver, CO 80210, USA
\and  Institute of Space and Astronautical Science, Japan Aerospace Exploration Agency, 3-1-1 Yosinodai, Chuo-ku, Sagamihara, Kanagawa, 252-5210, Japan
}
\authorrunning{Etxaluze et al.}

\abstract
{The Helix Nebula (NGC\,7293) is the closest planetary nebulae. Therefore, it is an ideal template for photochemical studies at small spatial scales in planetary nebulae.} 
{We aim to study the spatial distribution of the atomic and the molecular gas, and the structure of the photodissociation region along the western rims of the Helix Nebula as seen in the submillimeter range with {\it Herschel}.} 
{We use 5 SPIRE FTS pointing observations to make atomic and molecular spectral maps. We analyze the molecular gas by modeling the CO rotational lines using a non-local thermodynamic equilibrium (non-LTE) radiative transfer model.}
{For the first time, we have detected extended OH$^+$ emission in a planetary nebula. The spectra towards the Helix Nebula also show CO emission lines (from $J=4$ to 8), [N\,{\sc ii}] at 1461 GHz from ionized gas, and [C\,{\sc i}] ($^3\rm{P}_2-$$^3\rm{P}_1$), which together with the OH$^+$ lines, trace extended CO photodissociation regions along the rims. The estimated OH$^+$ column density is $\sim 10^{12}-10^{13}$ cm$^{-2}$. The CH$^+$ (1-0) line was not detected at the sensitivity of our observations. Non-LTE models of the CO excitation were used to constrain the average gas density ($n{\rm (H_2)}\sim (1-5)\times 10^5$ cm$^{-3}$) and the gas temperature ($T_{\rm k}\sim 20-40$ K).}
{The SPIRE spectral-maps suggest that CO arises from dense and shielded clumps in the western rims of the Helix Nebula whereas OH$^+$ and [C\,{\sc i}] lines trace the diffuse gas and the UV and X-ray illuminated clumps surface where molecules reform after CO photodissociation. [N\,{\sc ii}] traces a more diffuse ionized gas componnent in the interclump medium.}


\date{Received; accepted}
\titlerunning{{\it Herschel} spectral-mapping of the Helix Nebula (NGC 7293)}
\maketitle

\section{Introduction}
\indent\par {Planetary nebulae (PNe) represent the final stage in the evolution of low- and intermediate-mass stars like the Sun. These stars lose matter during the asymptotic giant branch (AGB) phase, and form an expanding circumstellar envelope. While the central stars evolve from AGB phase to planetary nebula (PN) phase, the effective temperatures of the stars rise, increasing the UV radiation. Molecules previously ejected in mass loss episodes from the progenitor star are photodissociated by the UV photons as the dissociation front advances through the gas. When the central star becomes hotter than $\sim 30,000$ K, the circumstellar gas is ionized. The nebular gas cools radiatively, i.e. by visible line emission. However, not all the envelope is ionized and neutral atoms and molecules dominate the gas cooling in photodissociation regions (PDRs) and in the shielded gas. In PNe, the neutral and molecular material form fragmented rims around the ionized nebula. Their infrared spectrum is characterized by [C\,{\sc i}], [O\,{\sc i}], [C\,{\sc ii}] atomic fine structure lines, and rotational lines of CO and H$_2$.  
}
\par{The Helix Nebula (NGC 7293) is the closest PN at a distance of $219\pm 30$ pc \citep{Harris07}, therefore it is a unique template to resolve the different structures expected in a planetary nebula (i.e. H\,{\sc ii} region, PDRs, and shielded molecular gas). The central star of the Helix Nebula is a white dwarf with an effective temperature $T_{\rm eff}\sim 120,000$ K, a luminosity of $L\sim 76$ $L_{\odot}$, and a mass of $M\sim 0.57$ $M_{\odot}$ \citep{Napiwotzki99, Traulsen05}. The central star is a source of X-rays at energies $\sim 0.25$ keV that arise from the stellar photosphere. In addition, {\it Chandra} and ROSAT observations revealed X-ray emission near $\sim 1$ keV \citep{Leahy94,Guerrero01}. Its origin remains uncertain.\\
}
\begin{table*}
\begin{center}
\caption{Observational parameters. \label{tab1}}
\renewcommand{\arraystretch}{1.2}
\begin{tabular}{lccccccc}
\hline\hline
ObsId$^a$ & Date&Target& Proposal & RA$_{\rm (J2000)}$$^b$& Dec$_{\rm (J2000)}$$^c$& Total time$^d$& v$_{\rm rad}$$^e$\\
&&&&&&(s) &(km s$^{-1})$ \\
\hline
1342256097 & 2012-11-25& T1&DDT\_mustdo\_7&$22^{\rm h}29^{\rm m}10.03^{\rm s}$ & $-20^{\rm o}48\arcmin 10.04\arcsec$ & 3612& -26.78\\
1342256098 &2012-11-25 &T2&DDT\_mustdo\_7&$22^{\rm h}29^{\rm m}22.84^{\rm s}$ & $-20^{\rm o}49\arcmin 18.79\arcsec$ & 3612&-26.77\\
1342256099& 2012-11-26&T3&DDT\_mustdo\_7&$22^{\rm h}29^{\rm m}22.84^{\rm s}$ & $-20^{\rm o}52\arcmin 30.81\arcsec$ & 3612&-26.78 \\
1342256100 & 2012-11-25&T4&DDT\_mustdo\_7&$22^{\rm h}29^{\rm m}10.03^{\rm s}$ & $-20^{\rm o}51\arcmin 22.00\arcsec$ & 3612&-26.79\\
1342257353 & 2012-12-17&T5 &OT2\_pvanhoof\_2 & $22^{\rm h}29^{\rm m}10.02^{\rm s}$ & $-20^{\rm o}49\arcmin 53.77\arcsec$ & 6992&-24.27\\
1342256744& 2012-12-08 & SPIRE 250 $\mu$m image & DDT\_mustdo\_7&$22^{\rm h}29^{\rm m}35.24^{\rm s}$ & $-20^{\rm o}50\arcmin 40.51\arcsec$ & 2047 & ...\\
\hline
\end{tabular}
\end{center}
\begin{list}{}{}
\item[$^a$]Observation identification number
\item[$^{b}$]Right ascension of the central detectors SLWC3 and SSWD4
\item[$^{c}$]Declination of the central detectors SLWC3 and SSWD4
\item[$^d$]Total integration time of each observation
\item[$^e$]Radial velocity of the {\it Herschel} telescope along the line of sight
\end{list}
\end{table*}
\begin{figure*}
\resizebox{\hsize}{!}{\includegraphics{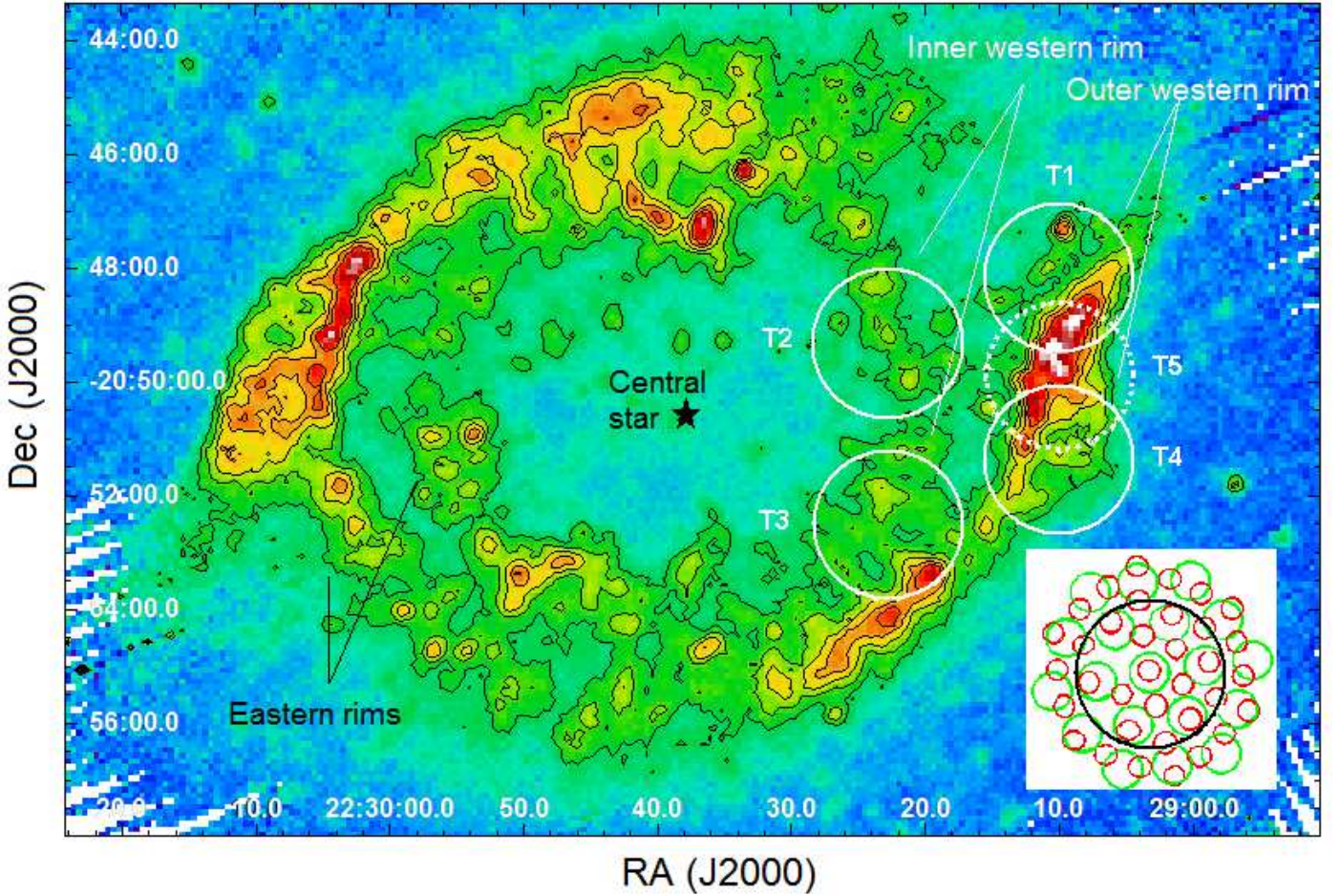}}
\caption{Footprint of the five SPIRE FTS observations over the SPIRE 250 $\mu$m photometric image. White circles show the layout of the unvignetted field of view (FoV$\sim$ 2.0 arcmin in diameter) of the four mustdo observations, and white-dashed circle shows the FoV of the pvanhoof observation. The total area coverage by the five observations is $\sim 6\times8$ arcmin$^2$. Layout of the FTS detector arrays (the SLW in green and the SSW in red) and the FoV (black) is shown in the lower right corner. The circle sizes of the detectors correspond to the FWHM of the beam. Black contours trace levels: 10, 12.5, 15, 17.5 and 20 MJy sr$^{-1}$} 
\label{fig1}
\end{figure*}

\par{Low-$J$'s CO and H$_2$ observations of the Helix Nebula show dense, neutral knots with ionized cometary tails located in the central region of the nebula \citep{Huggins02,ODell05,Hora06,Matsuura07}. \citet{Matsuura09} showed that it is in the inner region of the Helix, towards the central star, where tails are observed from the neutral globules probably created by the stellar winds from the central star \citep{Meaburn10,Matsuura09,Speck02}. The ionized cavity is surrounded by a double rim of dust, and molecular and atomic gas (see Figure~\ref{fig1}). The inner rim is a circular and fragmented ring around the central star \citep{Young99} with an inner radius $\sim 175$\arcsec ($\sim 0.2$ pc). The outer rim has an inner radius of $\sim 340\arcsec$ ($\sim 0.35$ pc). Knots at the envelope are difficult to resolve and seem to form clumps. The clumpiness and the strong UV fields in the envelope determine much of the physical and chemical conditions. 
}
\par{The Helix is probably an oxygen-rich planetary nebula (C/O= $0.87\pm 0.12$, \citet{Henry99}), with a high abundance of neutral carbon, $N$[C\,{\sc i}]$/N{\rm (CO)}\sim $ 6 measured towards the outer western rim \citep{Cox98}. The Helix molecular envelope is devoid of PAH's, which is consistent with its C/O ratio \citep{Hora06}. 
}
\par{So far, the lower energy levels of OH$^+$ emission lines have been detected in the ultra-luminous galaxies Mrk\,31 \citep{vanderWerf10} and NGC\,1068 \citep{Spinoglio12}. Most recently, \citet{vanderTak13} presented the first detection of extended OH$^+$ line emission in the galaxy, toward the Orion Bar, using the HIFI instrument onboard {\it Herschel}.
} 
\par{In this paper, we report the first detection of extended OH$^+$ emission in a circumstellar envelope, a key ion precursor for the oxygen chemistry. Simultaneously to this detection, \citet{Aleman13} also detected OH$^+$ emission in three planetary nebulae: NGC\,6445, NGC\,6720, and NGC\,6781 obtained in the {\it Herschel} Planetary Nebulae Survey (HerPlaNS) \citep{Ueta12}. Their work is also published in this volume. We also detected OH$^+$ in emission in NGC \,6853 planetary nebula. This nebula was also observed as part of the OT2 project (PI: P. van Hoof). The Helix Nebula and NGC\,6853 present very similar features in the SPIRE FTS spectra, showing emission lines of [N\,{\sc ii}], CO rotational lines (from $J=4-3$ to $J=8-7$), [C\,{\sc i}], and OH$^+$ along the molecular rims. A detailed analysis of the NGC\,6853 nebula will be presented in a forthcoming paper.
} 
\par{In this work, we study the physical and chemical conditions in the western rims of the Helix Nebula by analyzing the submm spectra from 447 GHz to 1550 GHz taken with the {\it Herschel} SPIRE Fourier Transform Spectrometer (FTS). The SPIRE FTS spectral maps allow us to study the molecular gas by modeling the CO rotational line emission and to study the distribution of the atomic and the molecular gas along the western rims. 
}
\section{Observation and Data Reduction}
 \indent\par{The SPIRE FTS instrument \citep{Griffin10}, on board the {\it Herschel} Space Observatory \citep{Pilbratt10}, observed the Helix Nebula with two detector arrays, the SPIRE Short Wavelength (SSW) Spectrometer with 37 detectors covering the frequency range $958-1546$ GHz and the SPIRE Long Wavelength (SLW) Spectrometer with 19 detectors covering the range $447-990$ GHz, with a pixel spacing of approximately twice the beam. Four sparse single pointing observations were made at a spectral resolution of 0.04 cm$^{-1}$ (1.2 GHz, equivalent to $300-940$ km\,s$^{-1}$) in bright mode through a Must-Do program with a total integration time of 3612 s for each observation. An additional single pointing observation was made through an Open Time 2 (OT2) project (PI: P. van Hoof) with a total integration time of 6992 s (see Table~\ref{tab1} for details).
}
\par{Figure~\ref{fig1} shows the layout of each pointing observation superimposed on the SPIRE 250 $\mu$m photometric image. The observations cover a total area of $\sim 6\times 8$ arcmin$^2$ tracing the inner and the outer rims at the western side of the Helix Nebula.
}
% Section 2, Subsection 1, Subsubsection 1, Paragraph 3
\par{The SPIRE FTS data were reduced with the {\it Herschel} Interactive Processing Environment (HIPE) version 11.1.0 \citep{Ott10}. The full width half maximum (FWHM) of the SPIRE FTS beam is wavelength dependent, changing between 17\arcsec and 21\arcsec for the SSW band and between 29\arcsec to 42\arcsec for the SLW band \citep{Makiwa13}. 
}
\par{The SPIRE 250 $\mu$m fully sampled image was processed with HIPE using standard version 10.0 build 2744. Additional to the standard processing the data were zero point corrected using the Planck dust model. The image is made with FWHM= 18.5", in MJy/sr, and at 6"/pixel. The SPIRE images will be discussed further in a forthcoming paper by Van de Steene et al. (2014), (in preparation). 
}
\par{We also analyze the H$_2$ distribution and its spatial correlation with the atomic and the molecular gas through the western rims of the Helix Nebula using a H$_2$ 2.122 $\mu$m image from \citet{Speck02}. The Helix was observed with the Near-Infrared Imager (NIRIM; \citet{Meixner99}) at the Mount Laguna 1 m telescope with the 2"/pixel scale (see \citet{Speck02} for the calibration details).  
}
\subsection{Spectral-maps with the SPIRE FTS}\label{Obs-map}
\indent\par{The five SPIRE FTS observations provide sparse spatial sampling of the molecular and ionized gas across the western rims of the Helix nebula. The integrated line intensities were obtained from the unapodized SPIRE FTS spectra after the subtraction of the continuum. To get a good accuracy on the central frequencies of spectral lines, the frequency scale was converted to the LSR frame, taking into account the radial velocity (${\rm v_{\rm rad}}$) of the {\it Herschel} telescope along the line of sight, so that the corrected central frequency was calculated as $\nu_{\rm corr.}=\nu (1- {\rm v_{\rm rad}}/{\rm c})$. The line shape was fitted with a classical $sinc$ function using the line-fitting tool available in HIPE. Sparsely sampled maps (Figure~\ref{fig3}) for several emission lines were made by interpolating the value of the integrated intensity of each line measured by adjacent detectors to halfway positions in order to analyze the spatial distribution of these species. The maps of those lines observed with the SLW detector have an angular resolution FWHM$\sim 30-35$\arcsec and maps of the lines in the frequency range of the SSW detector are spatially better sampled with an angular resolution FWHM$\sim 18$\arcsec. The total area covered by the five observations is $\sim 6\times8$ arcmin$^2$, and the resulting maps are centered at the position RA(J2000)= $22^{\rm h}29^{\rm m}16.3^{\rm s}$, Dec(J2000)= $-20^{\rm o}$50\arcmin 25.96\arcsec. The area was covered at 95 individual positions with detectors in the SLW array, and 185 positions with detectors in the SSW array. The maps were made with the same pixel size than the SPIRE 250 $\mu$m image, 6"/pixel. 
}
\begin{figure*}[t]
\resizebox{\hsize}{!}{\includegraphics{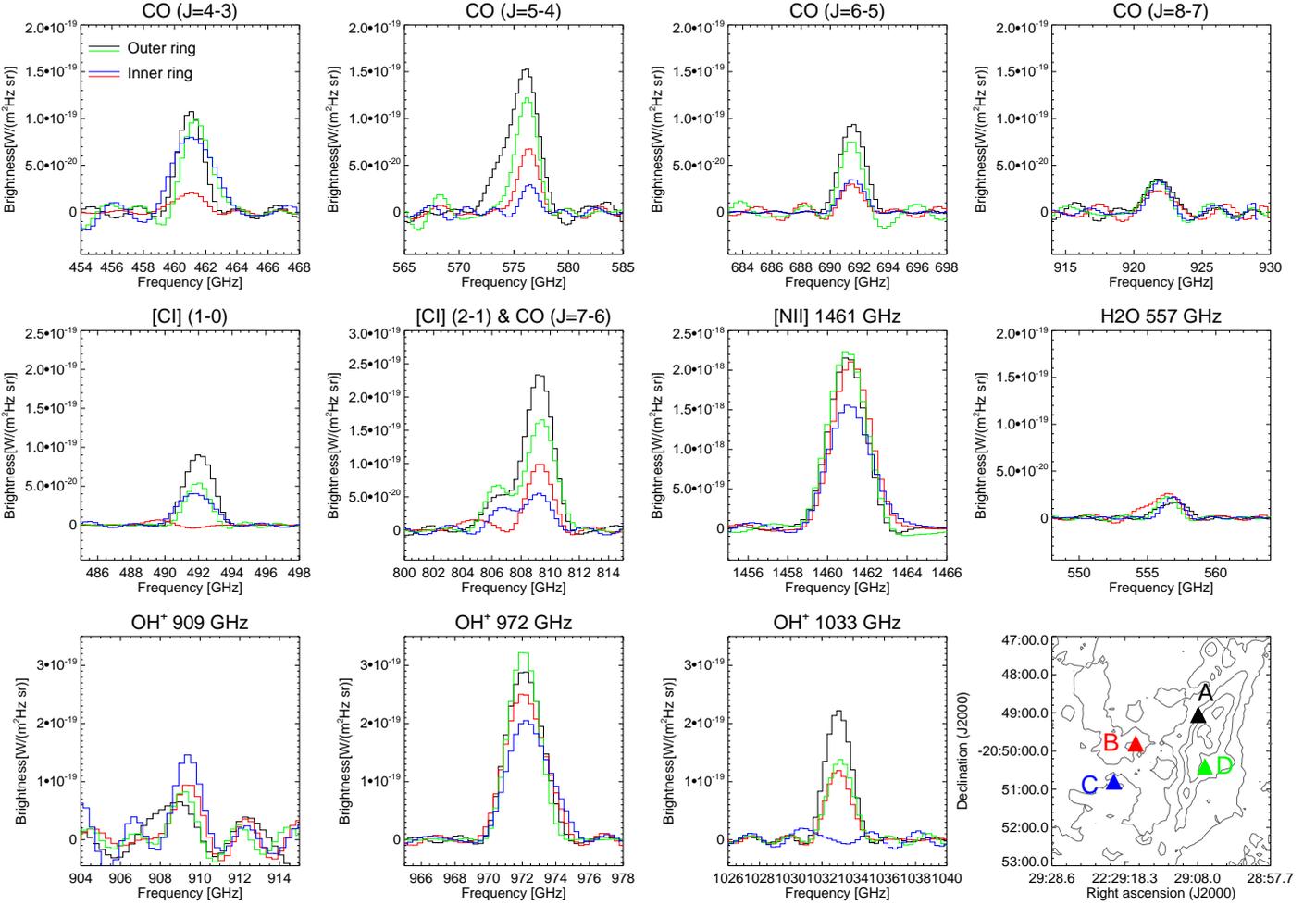}}
\caption{Line profiles of the atomic and the molecular lines observed with SPIRE-FTS in the western rims of the HELIX Nebula. In order to reduce the sidelobes, the spectra were apodized with an extended Norton-Beer function (FWHM= 15") to a resolution of 0.072 cm$^{-1}$. {\it Bottom-right}: The spectral lines are observed at four positions along the western rims, two positions in the inner rim (B in red, C in blue) and two position in the outer rim (A in black, D in green).}\label{fig2}
\end{figure*}
\begin{table*}
\begin{center}
\caption{Integrated line intensities for each species at each position indicated in Figure~\ref{fig2} (bottom-right). \label{tab2}}
\renewcommand{\arraystretch}{1.2}
\begin{tabular}{lcccccccc}
\hline\hline
Species & Transition&$\nu$&{$\Omega_{\rm beam}(\nu)$}&{Position-A}&{Position-B}&{Position-C}&{Position-D} \\
& &(GHz)&($10^{-8}$ sr) &($10^{-11}{\rm Wm^{-2}sr^{-1}}$)&($10^{-11}{\rm Wm^{-2}sr^{-1}}$)&($10^{-11}{\rm Wm^{-2}sr^{-1}}$)&($10^{-11}{\rm Wm^{-2}sr^{-1}}$)\\
\hline
CO & $J=4-3$ & 461.04 &14.2& 14.3$\pm$2.3&6.8$\pm$2.1&20.1$\pm$2.3&19.9$\pm$2.3\\
$\rm{[C\,I]}$& $^3{\rm P}_1$-$^3{\rm P}_0$ & 492.16&12.6&9.0$\pm$2.2&...&6.7$\pm$2.7&6.3$\pm$2.2\\
o-H$_2$O &  $1_{1,0}-1_{0,1}$ & 557.3 &10.0&&5.8$\pm$2.0&5.2$\pm$2.7&\\
CO & $J=5-4$ & 576.26 &9.54&47.9$\pm$2.2&23.4$\pm$1.0&13.3$\pm$2.2&45.0$\pm$2.6\\
CO & $J=6-5$ & 691.47 &8.06&29.4$\pm$2.2&7.5$\pm$2.2&11.0$\pm$2.2&26.5$\pm$2.2\\
CO & $J=7-6$ & 806.65 &8.59&12.3$\pm$2.2&7.1$\pm$2.7&&13.0$\pm$2.2\\
$\rm{[C\,I]}$ & $^3{\rm P}_2$-$^3{\rm P}_1$ & 809.34 &8.62&58.3$\pm$2.2&25.4$\pm$2.0&8.3$\pm$2.7&12.5$\pm$2.2\\
OH$^+$ & $1_0-0_1$ & 909.27 &9.77&8.5$\pm$2.2&10.7$\pm$2.0&19.1$\pm$2.7&9.0$\pm$2.7\\
CO & $J=8-7$ & 921.79 & 9.92&7.4$\pm$2.2&6.5$\pm$2.2&4.7$\pm$2.7&10.9$\pm$2.2\\
OH$^+$ & $1_2-0_1$ & 971.8 &2.81&70.1$\pm$2.2&40.7$\pm$2.0&43.3$\pm$2.8&92.8$\pm$2.2\\
OH$^+$ & $1_1-0_0$ & 1033.23 &2.51&45.3$\pm$8.5&39.1$\pm$9.2&...&35.2$\pm$9.7\\
$\rm{[N\,II]}$ & $^3{\rm P}_1$-$^3{\rm P}_0$ & 1461.13 &2.21&506.9$\pm$8.5&498.8$\pm$9.2&396.7$\pm$9.2&539.6$\pm$9.7\\
\hline
\end{tabular}
\end{center}
\end{table*}

\section{Results}
\indent\par{Figure~\ref{fig2} shows the spectral lines detected at four different positions, two in the inner western rim and the other two in the outer western rim. The spectra show CO rotational lines in emission from $J=4-3$ to $J=8-7$ transitions, and rotational lines in emission from the lowest energy levels of OH$^+$ at 909.1 GHz, 971.9 GHz, and 1033 GHz. The detection of extended OH$^+$ lines in emission is particularly interesting. The OH$^+$ critical densities for collisional excitation are high, $n_{\rm cr}> 10^6$ cm$^{-3}$, therefore, most of the OH$^+$ detections to date have been observed in absorption against the strong submm background continuum emission (i.e., toward the Sgr B2 molecular cloud, Sgr~A$^*$, in the Orion BN/KL outflow, and toward several low and high mass star forming regions \citep{Gerin10,Neufeld10,Wyrowski10,Gupta10,Etxaluze13,Goicoechea13}). 
}
\par{The spectra also show the fine structure emission lines of [C\,{\sc i}] ($^3\rm{P}_1-$$^3\rm{P}_0$) at 492.39 GHz, [C\,{\sc i}] ($^3\rm{P}_2-$$^3\rm{P}_1$) at 809.34 GHz, which is blended with the CO $J=7-6$ line at 806.65 GHz, and [N\,{\sc ii}] ($^3\rm{P}_1-$$^3\rm{P}_0$) at 1461 GHz. Most lines are brighter at the two positions observed in the outer western rim, except [N\,{\sc ii}], which seems equally bright in the inner and in the outer rims. [N\,{\sc ii}] is the brightest line in the SPIRE-FTS spectra. The spectra also show a weak emission line from the ground-state ortho-H$_2$O at 557.3 GHz. 
}
\par{Table~\ref{tab2} summarizes the line surface brightness of all the observed lines between 430 and 1550 GHz, obtained from the SPIRE FTS unapodized spectra at four different positions toward western rims. The Helix Nebula presents a very complex velocity structure when observed at much higher spectral resolution. \citet{Zack13} observed CO rotational lines (from $J=1-0$ to $J=3-2$) toward the Helix with a line spectral resolution $\sim 0.9-1.3$ km\,s$^{-1}$, revealing multiple velocity components in the three CO transitions. The observed velocities are in the range $\sim -(11-30)$ km\,s$^{-1}$ toward the western rims. The CO rotational lines ($J\ge 4$) observed with the SPIRE FTS may consist of several blended velocity components unresolved due to the lower resolution ($\ge 300$ km\,s$^{-1}$), preventing a detailed analysis of the complex kinetic structure of the Helix Nebula. 
}
\subsection{Spatial distribution of [N\,{\sc ii}], C, CO, and OH$^+$}
\indent\par
\begin{figure*}[t]
\resizebox{\hsize}{!}{\includegraphics{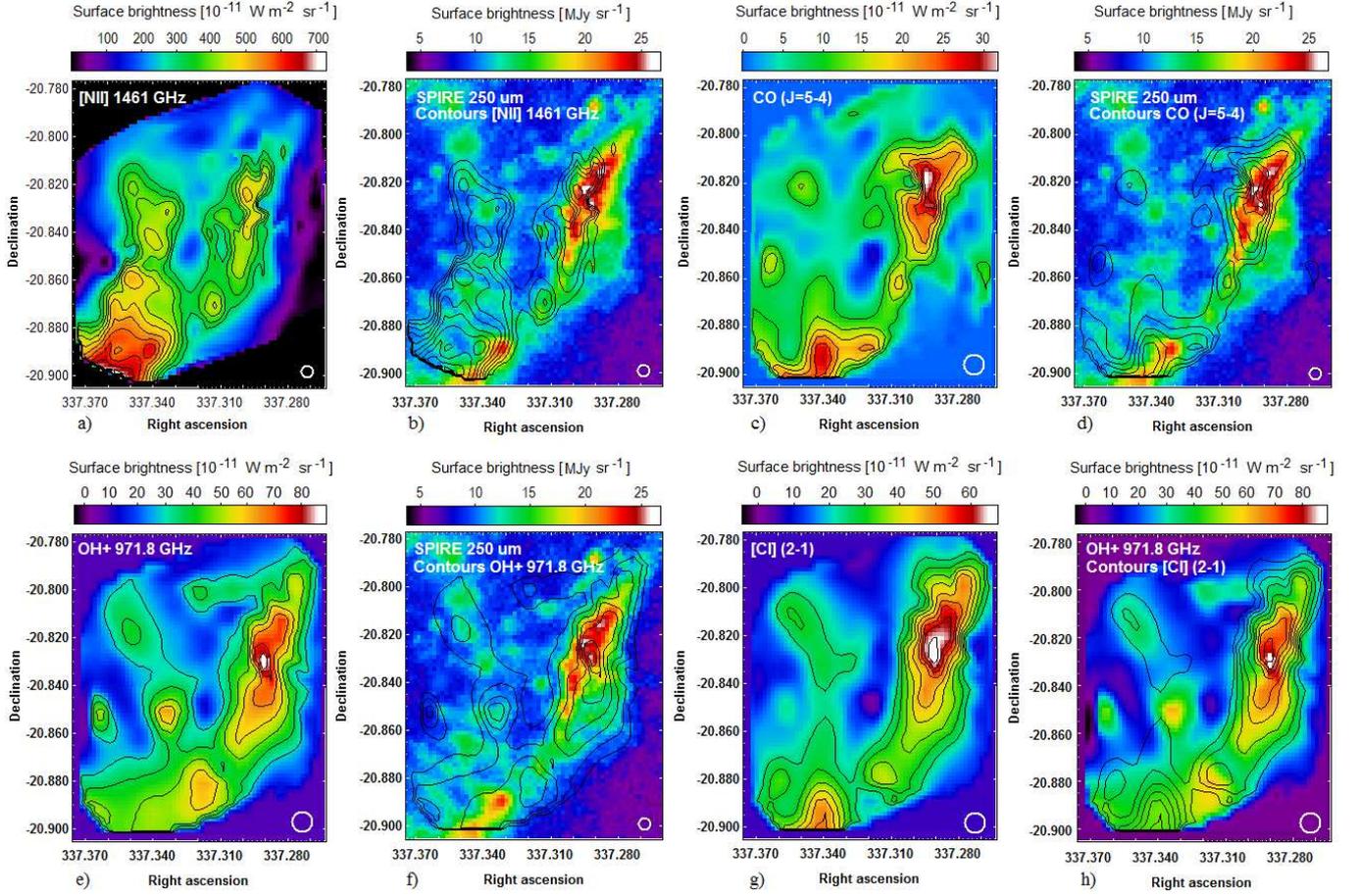}}
\caption{a), c) and e) Surface brightness maps of [N\,{\sc ii}] at 1461 GHz, CO ($J= 5-4$) and OH$^+$ at 971.8 GHz, respectively. The contours of these maps are overlaid on the SPIRE 250 $\mu$m image: b), d) and f), respectively. g) shows the image of the surface brightness of [C\,{\sc i}] (2-1). h) shows the contours of the [C\,{\sc i}] (2-1) intensity distribution overplotted on the OH$^+$ 971.8 GHz map. All the maps are centered at RA(J2000)= $22^{\rm h}29^{\rm m}16.3^{\rm s}$, Dec(J2000)= $-20^{\rm o}$50\arcmin 25.96\arcsec. The maps are resampled with a scale of 6"/pixel.} 
\label{fig3}
\end{figure*}

{Figure~\ref{fig3} shows the integrated line intensity sparsely sampled maps of [N\,{\sc ii}] ($^3\rm{P}_1-$$^3\rm{P}_0$) at 1461 GHz, along with the CO $J= 5-4$ rotational line, OH$^+$ ($N= 1-0$) at 971.9 GHz and [C\,{\sc i}] ($^3\rm{P}_2-$$^3\rm{P}_1$) at 809.34 GHz, and their contours superimposed on the SPIRE 250 $\mu$m image. Among the detected CO rotational lines, the $J=5-4$ line is the brightest along the molecular rims. This line was chosen to trace the CO distribution along the western rims. The comparison of these maps allows us to study the excitation mechanisms of the emission lines and the distribution of the different species through the region. The maps cover a total area of $\sim 6\times 8$ arcmin$^2$, corresponding to $\sim 0.4\times 0.5$ pc$^2$ for a distance of 219 pc \citep{Harris07}. 
}
\par{Figure~\ref{fig3}\,a) shows the distribution of the [N\,{\sc ii}] ($^3\rm{P}_1-$$^3\rm{P}_0$) surface brightness along the western edge of the Helix Nebula, and Figure~\ref{fig3}\,b) shows a contour plot of the [N\,{\sc ii}] intensity distribution superimposed on the SPIRE photometric image at 250 $\mu$m. The 250 $\mu$m image traces the distribution of cold dust through the Helix Nebula. The SPIRE 250 $\mu$m image (Figure~\ref{fig3}\,b), d) and f)) shows cold dust distributed through the inner and the outer western rims with a bright structure of cold dust in the outer western rim. The [N\,{\sc ii}] line traces gas ionized by radiation from the central star, which is located at a distance of $\sim 241$ arcsec ($\sim 0.25$ pc) to the east of the inner western rim. The [N\,{\sc ii}] line traces the position of the two rims. The inner western rim observed in the [N\,{\sc ii}] map matches the position of the inner western rim traced by the cold dust on the SPIRE 250 $\mu$m image. In the outer western rim, the [N\,{\sc ii}] emission extends along the face pointing toward the central star, along the inner side of the outer western rim. 
}
\begin{table*}
\centering
\caption{$T_{\rm rot}$ (K), $N$(CO) (cm$^{-2}$), and $n\rm{(H_2)}$ (cm$^{-3}$) throughout the outer western rim of the Helix Nebula. $T_{\rm{rot}}\sim T_{\rm k}$.}
\label{tab3}
\renewcommand{\arraystretch}{1.2}
\begin{tabular}{ccccccc}
\hline\hline
   {Detector}&{R.A.}& {Dec.}& $T_{\rm rot}$ &{$N$(CO)}& {$T_{\rm k}$} &$n\rm{(H_2)}$   \\
   {SLW} &{(J2000)} &{(J2000)}&{(K)}&{($10^{14}$cm$^{-2}$)} &{(K)}&{($10^{5}$cm$^{-3}$)}\\  
\hline  
 B2 & 22:29:13.5 & -20:49:57.8 & 29$\pm$3 &10.0$\pm$1 & 38 & 1.0  \\
 C2 & 22:29:11.1 & -20:50:41.8 & 28$\pm$2&13$\pm$1 & 31 & 5.0 \\
 C3 & 22:29:10.2 & -20:49:48.8 & 32$\pm$2 &12.0$\pm$1 & 33 & 6.0 \\
 C4 & 22:29:09.0 & -20:48:58.3 & 29$\pm$1 &11.2$\pm$0.7 & 30  & 6.0 \\
 C5 & 22:29:07.7 & -20:48:16.5 & 28$\pm$3 &7$\pm$1 & 29 &  4.0 \\
 D1 & 22:29:08.7 & -20:51:22.3 & 25$\pm$3 &8$\pm$1 & 26  & 3.0  \\
 D3 & 22:29:06.8 & -20:49:41.6 & 22$\pm$2 &8$\pm$1 & 23  & 3.0 \\
\hline
\end{tabular}
\end{table*}
 
\par{Figures~\ref{fig3}\,c) and~\ref{fig3}\,d) show the distribution of the CO $J=5-4$ emission line throughout the Helix, and the CO $J=5-4$ surface brightness contours overlaid on the SPIRE 250 $\mu$m image, respectively. The CO $J=5-4$ line is mainly distributed along the outer western rim tracing the dust emission at 250 $\mu$m. The CO emission peak is coincident with the strongest dust continuum emission in Figure~\ref{fig3}\,d). The inner western rim is less evident.
}
\par{The OH$^+$ 971.8 GHz and [C\,{\sc i}] ($^3\rm{P}_2-$$^3\rm{P}_1$) 809.34 GHz surface brightness distributions (Figure~\ref{fig3}\,e) and g), respectively) trace the inner and the outer western rims on the SPIRE 250 $\mu$m image (Figure~\ref{fig3}\,f)), with the outer rim being brighter than the inner. The OH$^+$ and [C\,{\sc i}] distributions are spatially coincident peaking at the same position in the outer western rim (Figure~\ref{fig3}\,h)). The OH$^+$ and [C\,{\sc i}] emission lines are brighter than the CO $J= 5-4$ line and their brightness distributions extend to the west of the outer rim, while the CO $J= 5-4$ line emission distribution is delimited along the center of the rim. 
}
\par
\begin{figure}
\resizebox{\hsize}{!}{\includegraphics{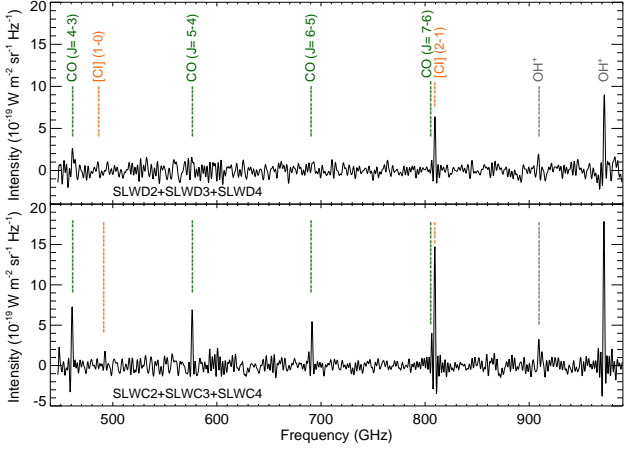}}
\caption{Lower: Co-added SPIRE FTS spectra of three SLW detectors covering the center of the outer western rim. Upper: Co-added SPIRE FTS spectra of three SLW detectors covering the west side of the outer western rim.} 
\label{fig4}
\end{figure}

{Figure~\ref{fig4} shows two spectra, each of them made by co-adding the spectra measured by three SLW detectors at three different positions. Co-adding the spectra, we achieve a better signal-to-noise ratio to identify the intrinsically weak atomic and the molecular emission lines in the region. The lower spectrum of Figure~\ref{fig4} is as result of co-adding the signal from detectors SLWC2, SLWC3, and SLWC4 (see Figure~\ref{fig5}\,a)), which cover the emission along the outer western rim. Detectors SLWD2, SLWD3, and SLWD4 (Figure~\ref{fig5}\,a)) cover the gas emission at the west side of the outer western rim and their signals were also co-added (Figure~\ref{fig4} top). The final co-added spectra show strong emission lines of OH$^+$ at 909.3 GHz and 971.8 GHz, and [C\,{\sc i}] at 492.16 GHz and 809.34 GHz, together with CO rotational lines from $J= 4-3$ to $J= 7-6$ along the outer western rim. The CO rotational lines are below the detection limit along the west side of the outer western rim, and only the OH$^+$ and the [C\,{\sc i}] lines are bright enough to be detected. 
}
\subsection{CO Rotational Ladder}
\indent\par{We plotted the CO rotational population diagrams using the CO lines from $J= 4-3$ to $J= 8-7$ at seven different positions through the outer western rim (Figures~\ref{fig5}\,a) and b)). For detailed analysis, we use target T5, as the integration time is longest and the quality of its spectrum is the best. The other four observations were not used as they had lower integration times, with consequently poorer signal to noise ratios, and only three CO lines were detected at each position. The CO rotational diagrams provide the rotational temperature, $T_{\rm rot}$. Assuming extended and optically thin emission, $T_{\rm rot}$ is a lower limit to the gas kinetic temperature, $T_{\rm k}$. We can also derive the beam-averaged CO column density, $N{\rm (CO)}$.
}
\par{Figure~\ref{fig5}\,b) shows the CO rotational diagrams at seven different positions throughout the outer western rim. The rotational diagrams are best fitted with a single excitation temperature component. The rotational temperatures are relatively low, over the range $\sim$20-30 K. The rotational temperature along the outer rim is $\sim$30 K, from north to south. The temperatures are lower on the west side of the outer western rim. The CO column densities are in the range $\sim$(7-12)$\times 10^{14}$ cm$^{-2}$ with the highest values distributed along the center of the outer western rim. The values are in good agreement with those found by \citet{Zack13} from lower-$J$ CO observations at the position of the western rims who determined $N\rm{(CO)}$ values of $\sim$7$\times 10^{14}$ cm$^{-2}$ and $\sim$14$\times 10^{14}$ cm$^{-2}$, for two different velocities components at -11 and -21 km\,s$^{-1}$, respectively. The values of the rotational temperature and the CO column density at each position are indicated on Table~\ref{tab3}.
}
\subsection{Modeling the CO rotational ladder}
\indent\par
\begin{figure}[]
        \centering
        \begin{subfigure}[]
                \centering
 \resizebox{\hsize}{!}{\includegraphics{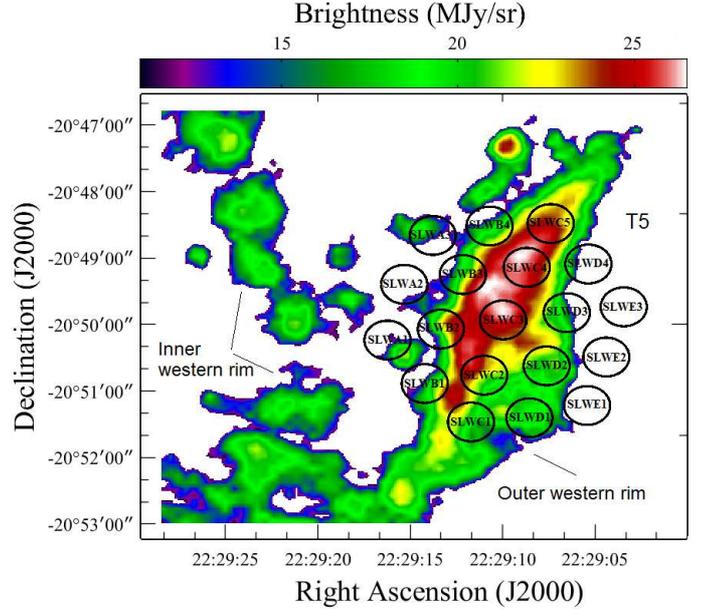}}
                       \end{subfigure}

        \begin{subfigure}[]
                \centering
  \resizebox{\hsize}{!}{\includegraphics{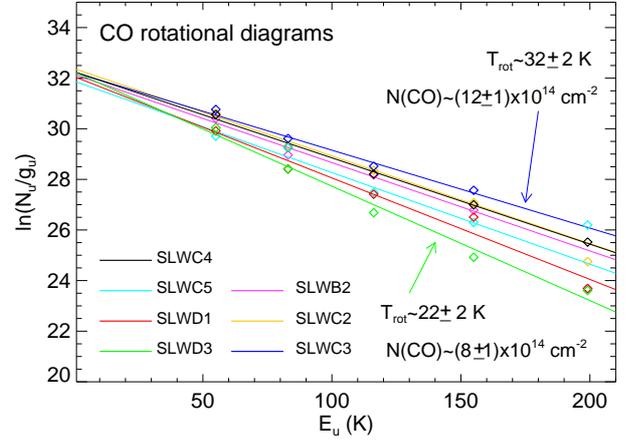}}
                        \end{subfigure}

        \begin{subfigure}[]
                \centering
   \resizebox{\hsize}{!}{\includegraphics{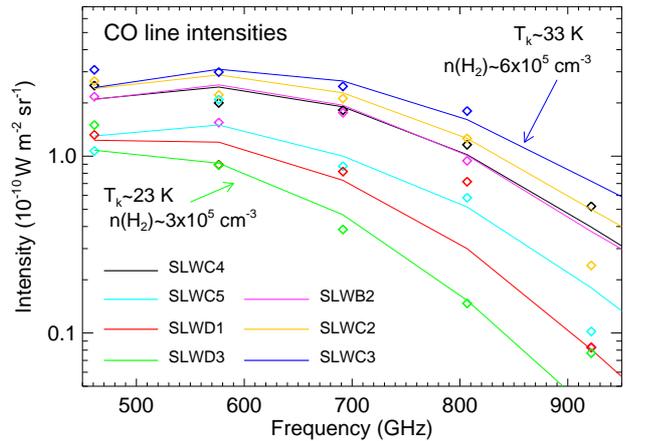}}
                      \end{subfigure}
        \caption{(a) Footprint of the SLW array (target T5) over the outer western rim. (b) CO rotational diagrams showing the column density per statistical weight versus the energy of the upper level. The CO rotational diagrams are measured at seven different positions along the outer western rim (SLWC2, SLWC3, SLWC4, SLWC5, SLWB2, SLWD1 and SLWD3). (c) Integrated CO intensities fitted with non-LTE models at the same seven positions previously mentioned.} \label{fig5}
\end{figure}

{Observations of CN, CO, and HCO$^+$ of the Helix nebula, by \citet{Bachiller97}, show relatively high gas densities, $n{\rm (H_2)}\sim (1-4)\times 10^5$ cm$^{-3}$. This density is below or comparable to the critical densities for collisional excitation of the observed CO lines. Therefore, we assume the CO excitation to be in a regime where $T_{\rm rot}\lesssim T_{\rm k}$. In order to constrain the physical conditions that reproduce the observed CO intensities towards the outer western rim, we treated the non-LTE excitation and radiative transfer problem with MADEX, an LVG (large velocity gradient) code \citep{Cernicharo12}. As a first approximation, the CO column density is assumed to be similar to that determined from the CO rotational diagrams. Thus, we searched for a combination of the kinetic temperature, $T_{\rm k}$, and the molecular hydrogen density, $n{\rm(H_2)}$, that better reproduces the observed CO intensities from $J= 4-3$ to $J= 8-7$ (Figure~\ref{fig5}\,c)). A line width of $\Delta v=8$ km\,s$^{-1}$ was adopted, similar to that measured on the CO $J= 3-2$ line by \citet{Zack13}. The best fit is obtained for kinetic temperatures $T_{\rm k}\sim 20-40$ K across the Helix, with $T_{\rm k}\sim 40$ K in the inner side of the outer rim and decreasing outwards. The gas density is $n{\rm (H_2)}\sim (1-6)\times 10^5$ cm$^{-3}$. These results agree with previous studies, such as those obtained by \citep{Bachiller97}. Gas densities in the Helix were estimated to be $n{\rm (H_2)}\sim 10^4-10^5$ cm$^{-3}$ based on observed transitions of H$_2$ \citep{Meixner05}. \citet{Zack13} estimated gas kinetic temperatures $T_{\rm k}\sim 15-40$ K and general densities $n{\rm (H_2)}\sim (0.1-5)\times 10^5$ cm$^{-3}$ in the Helix based on five lines of H$_2$CO ($J_{\rm Ka,Kc}= 1_{0,1}\rightarrow 0_{0,0}$, $2_{1,2}\rightarrow 1_{1,1}$, $2_{0,2}\rightarrow 1_{0,1}$, $2_{1,1}\rightarrow 1_{1,0}$, and $3_{0,3}\rightarrow 2_{0,2}$) and three CO rotational lines ($J=1-0$, $J=2-1$ and $J=3-2$). However, the gas traced by the low-$J$'s ($J\le 3$) at the position of the western rims by \citet{Zack13}, is associated with a colder gas component, $T_{\rm k}\sim 15-20$ K, of lower densities $n{\rm (H_2)}\sim (0.3-1.4)\times 10^5$ cm$^{-3}$.
}
\par{The average fractional abundance of CO relative to molecular hydrogen in the Helix is ${\rm \chi_{(CO)}}\sim 2.2\times 10^{-4}$ \citep{Zack13} therefore, the molecular hydrogen column density traced by the mid-$J$ CO lines is in the range $N{\rm (H_2)}\sim (2-6)\times 10^{18}$ cm$^{-2}$. 
}

\subsection{OH$^+$ Local thermodynamic equilibrium analysis}
\indent\par{The rates of collisional excitation of OH$^+$ with H and H$_2$ are unknown. The OH$^+$ lines have high Einstein coefficients $A_{\rm 10}> 10^{-2}$ s$^{-1}$, three orders of magnitude higher than those of the mid-$J$ CO lines, and thus the critical densities of the OH$^+$ ($N=1-0$) are higher than the densities estimated by modeling the CO rotational ladder in non-LTE, along the molecular rings in the Helix. As a first approximation, the OH$^+$ column density were estimated assuming a Boltzmann distribution of the rotational level population with an excitation temperature of $T_{\rm ex}= 10-20$ K. For $T_{\rm ex}= 20$ K, we estimated a column density of $N{\rm (OH^+)}\sim 7\times 10^{11}$ cm$^{-2}$ along the outer western rim, and increases by a factor of ten for $T_{\rm ex}= 10$ K.   
}
\par{The estimated OH$^+$ abundance relative to H$_2$ is $\sim 10^{-8}-10^{-7}$. Similar OH$^+$ abundances were inferred in the Orion Bar by \citet{vanderTak13} based on Meudon PDR models, assuming that most of the OH$^+$ emission arises from the outermost UV illuminated layers of the PDR at extinctions $A_V<0.4$.  
}
\section{Discussion}
\indent\par{We have presented the first detection of extended OH$^+$ emission in a circumstellar envelope. The central star of the Helix Nebula with a temperature $T_{\rm eff}\sim 120,000$ K \citep{Napiwotzki99, Traulsen05} is a strong source of ionizing UV radiation. Previous studies show the central region of the Helix ($\sim 0.25$ pc in radius) dominated by highly ionized gas such as He\,{\sc ii} \citep{ODell98} and [O\,{\sc iv}] \citep{Leene87}. Surrounding this highly ionized region, there are the inner and outer rims. Figure~\ref{fig1} presents the rims as a fragmented and clumpy structure traced by the dust continuum emission at 250 $\mu$m. The rims are subjected to the UV radiation from the central star, which enhances the rate of ionization in the clumps \citep{Ali01}. 
}
\subsection{Ionic, atomic and molecular spatial distribution}
\par
\begin{figure*}[t]
\resizebox{\hsize}{!}{\includegraphics{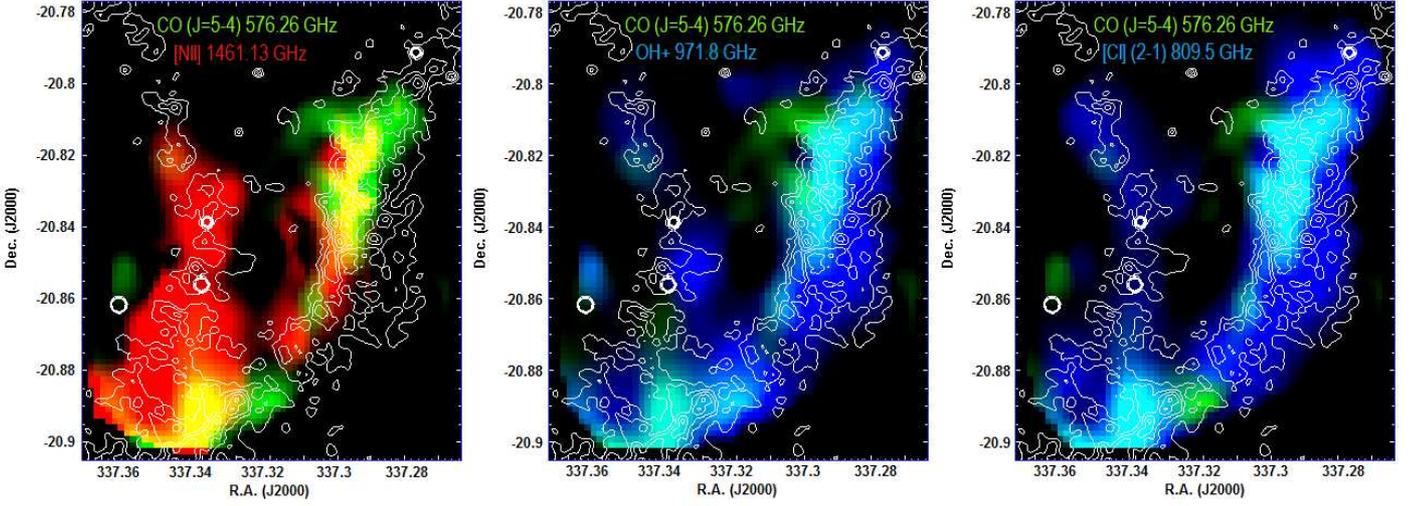}}
\caption{Three composite images showing the distribution of different atoms, ions, and molecules along the western rims of the Helix Nebula. a) [N\,{\sc ii}] at 1461 GHz in red, CO ($J=5-4$) in green. Yellow areas are the regions where [N\,{\sc ii}] emission coincide with CO ($J=5-4$) emission. b) CO ($J=5-4$) in green and OH$^+$ in blue. c) CO ($J=5-4$) in green and [C\,{\sc i}] ($^3{\rm P}_2$-$^3{\rm P}_1$) in blue. Cyan areas are the regions where CO ($J=5-4$) emission coincide with the OH$^+$ emission in b), and with the [C\,{\sc i}] emission in c). White contours trace the H$_2$ distribution on the H$_2$ 2.122 $\mu$m map.} 
\label{fig6}
\end{figure*}

{Figure~\ref{fig6}\,a) shows a composite image of [N\,{\sc ii}] 1461 GHz (red), and the CO $J=5-4$ rotational line (green), together with the H$_2$ contours (white). [N\,{\sc ii}], with an ionization potential of 14.53 eV, traces the low-excitation ionized gas. [N\,{\sc ii}] follows the molecular H$_2$ emission along the inner western rim and the diffuse gas along the east side of the outer western rim that faces the central star. The [N\,{\sc ii}] intensity drops drastically below the SPIRE FTS detection limit to the west from the center of the outer western rim. \citet{Matsuura09} studied the Helix rims, using images in the 2.122 $\mu$m H$_2$ $\upsilon=1\rightarrow 0$ $S(1)$ line and [N\,{\sc ii}] at 658.4 nm. They found that H$_2$ is associated only with high-density clumps while [N\,{\sc ii}] is distributed through the diffuse gas between them. 
}
\par{The CO $J=5-4$ line emission is mainly located along the outer western rim. Throughout the outer rim, the CO column density is $N{\rm(CO)}\sim 12\times 10^{14}$ cm$^{-2}$. The estimated molecular hydrogen density is $n{\rm (H_2)}\sim6\times 10^5$ cm$^{-3}$. $N{\rm (CO)}$ and $n{\rm (H_2)}$ decrease approximately half to the west of the outer rim. Maps of high density tracers such as HCO$^+$ $J=1-0$ \citep{Zeigler13} and CO $J=2-1$ \citep{Young99} also present a clear intensity maximum in the outer rim. The map of the CO $J=2-1$ line with a critical density $n_{\rm cr}\sim 10^4$ cm$^{-3}$, shows a clumpy and fragmented inner western ring while the CO $J=5-4$ line ($n_{\rm cr}\sim 10^5$ cm$^{-3}$) is barely detected in the SPIRE FTS spectra along the inner western rim.
}
\par{Figure~\ref{fig6}\,b) and c) show the distribution of OH$^+$ and [C\,{\sc i}] ($^3{\rm P}_2$-$^3{\rm P}_1$), respectively, together with the CO $J=5-4$ line and the H$_2$ intensity contours. As presented on Figure~\ref{fig3}\,h), the distributions of OH$^+$ and [C\,{\sc i}] are spatially well correlated. While the CO distribution is confined in the east side and the center of the outer rim, the OH$^+$ and [C\,{\sc i}] distributions extend to the west side of the outer western rim. OH$^+$ and [C\,{\sc i}] also trace the inner rim. OH$^+$ and [C\,{\sc i}] distributions are well correlated with the molecular hydrogen H$_2$ 2.122 $\mu$m line distribution through the entire region covered by the SPIRE FTS. This suggests that the extended OH$^+$ and [C\,{\sc i}] emission arises from the illuminated surface of the molecular clumps. 
}
\par{According to PDRs models \citep{Hollenbach99}, the [C\,{\sc i}] emission layer traces the illuminated cloud surface i.e., the east side of the outer western rim that faces the exciting star. However, this structure is not seen in our maps of the [C\,{\sc i}] and CO ($J=5-4$) distributions in Figure~\ref{fig6} c). Instead, the distribution of the [C\,{\sc i}] emission coincides with the CO emission along the center of the outer rim, but extends further to the west side of the outer rim, correlating spatially with the H$_2$ and OH$^+$ distributions. \citet{Young99} observed the Helix Nebula in the CO $J=2-1$ line with a velocity resolution of 1.5 km\,s$^{-1}$, revealing a complex velocity structure and suggesting that the nebula consists on an expanding equatorial ring with arcs and filaments extending around the ionized cavity. \citet{ODell04} suggested that the Helix is composed of an inner disk surrounded by an outer ring inclined 23$^{\rm o}$ to the plane of the sky. Recently, \citet{Zeigler13} mapped the Helix at the $J=1-0$ transition of HCO$^+$ at 89 GHz at high spectral resolution ( 1.68 km\,s$^{-1}$). The HCO$^+$ map reveals the Helix as a barrel-like structure composed of two bipolar outflows inclined 10$^{\rm o}$ to east relative to the line of sight. The bi-polar structure was previously suggested by \citet{Meaburn05}. The high velocity material forms blue and red-shifted disks. The inner western rim is part of the blue-shifted disk and the outer western rim is part of the red shifted disk. The discrepancy observed between the [C\,{\sc i}] and CO $J=5-4$ distributions and the predicted structure on PDR models, as well as the particular atomic and molecular stratification observed toward both western rims might be caused by projection effects due to the tilted barrel-like shape and the fact that the western rims are not co-located in the plane of the sky but separated at least 0.3 pc.
}
\par{The observed spatial distributions of CO and [C\,{\sc i}] requires a clumpy environment within a nebula. Clumps have relatively higher optical depths, shielding UV radiation and protecting molecules in the inner layers. UV radiation can penetrate further between clumps \citep{Spaans96,Kramer08}, which is filled with ionized gas. For molecular gas in clumps within the ionized region of a nebula, PDRs are formed at the surface of the clumps, dissociating the CO molecules on the surfaces of clumps facing the central star, and also throughout the diffuse gas between the clumps \citep{Speck02}. Therefore, the CO emission likely arises from the highest density gas component in the clumps.}
\subsection{CO photodissociation and OH$^+$ formation}
\par{The entire Helix Nebula contains a rich variety of molecules \citep{Young99,Tenenbaum09,Zack13}. In the rims, the abundance of CO is likely regulated by photodissociation by UV photons. Molecules are expected to be destroyed in the photodissociation fronts. The FUV (far-ultraviolet) field passing through the inner western rim dissociate the CO molecules. Once CO is photodissociated, oxygen can form OH$^+$ via
}
\begin{center}
O + H$^+$ $\rightleftarrows$ O$^+$ + H 

O$^+$ + H$_2$$\rightarrow$ H + OH$^+$ 
\end{center}

\par{The formation of H$^+$ via cosmic ray ionizations or X-rays, is followed by recombination of H$^+$ with electrons and with neutral or negatively charged polycyclic aromatic hydrocarbons (see detailed models in \citet{Hollenbach12}). The fact that the Helix Nebula is devoid of PAHs \citep{Hora06} likely enhance the formation of OH$^+$ through the western rims. The strong X-rays emission at energies $\sim$ 1 keV observed in the Helix Nebula \citep{Leahy94,Guerrero01}, likely increases the hydrogen ionization rates and this may enhance the formation of OH$^+$. However, NGC\,6853 doesn't show such a strong X-ray emission. NGC\,6853 presents low X-rays consistent with photospheric emission at energies $\sim$ 0.2 keV \citep{Kastner12} similar to the low X-ray emission observed in the Helix Nebula. The spectrum of NGC\,6853 also shows OH$^+$ lines in emission (forthcoming paper), suggesting that the strong X-ray emission observed in the Helix Nebula, may not be relevant on the formation of OH$^+$.     
}
\par{CO is photodissociated through the inner rim. This would explain the weak CO emission detected almost below the SPIRE FTS detection limit, and the good correlation between the [C\,{\sc i}] and OH$^+$ emission observed along the inner rim where CO molecules are photodissociated. 
}
\par{The UV radiation reaching the outer western rim penetrates between the clumps, photodissociating the CO molecules throughout the entire rim. The brightest CO emission is spread along the center of the outer western rim. In this region, the CO photodissociation is reduced by shielding by H$_2$, dust, and shelf-shielding. The dust continuum emission and $N{\rm (H_2)}$ peak along the center of the outer western rim, where $N{\rm (CO)}\sim 1.2\times 10^{15}$ cm$^{-2}$. CO photodissociation is strongly reduced at CO column densities higher than $10^{15}$ cm$^{-2}$ \citep{Bally82,Lee96}. Beyond the center region of the outer western rim, the molecular hydrogen and CO column densities are lower than in the center of the rim. CO photodissociation is more efficient and the CO intensities again drop below the SPIRE FTS detection limit, and only [C\,{\sc i}] and OH$^+$ are detected.   
}
\par{Once OH$^+$ is formed, it is destroyed by reactions with electrons and H$_2$ \citep{Hollenbach12}. Depending on their relative contribution, the oxygen chemistry will proceed and ultimately form water vapor:
} 
\begin{center}
OH$^+$ + H$_2$$\rightarrow$ H$_2$O$^+$+ H$_2$$\rightarrow$ H$_3$O$^+$ + $\mathrm{e^-}$ $\rightarrow$ H$_2$O 
\end{center}
\par{The non-detection of H$_2$O$^+$ and H$_3$O$^+$ suggests that the OH$^+$ emission traces low-molecular fraction gas through the western rims of the Helix Nebula \citep{Neufeld10}. 
}
\par{The ortho-H$_2$O line at 557.3 GHz is also detected (see Figure~\ref{fig2}) although this line is much weaker than the OH$^+$ lines. The fact that only the H$_2$O ground-state transition is detected suggests that water vapor emission is likely arising from cooler shielded clumps.   
}
\subsection{Carbon column density and CH$^+$ non-detection}
\par{The fine structure [C\,{\sc i}] lines are one of the best diagnostic to probe the PDRs structures. It is the major form of carbon in the neutral gas of the Helix and contributes to the cooling of the gas. The [C\,{\sc i}] column densities estimated with MADEX are $N{\rm (C\,I)}\sim 6.5\times 10^{15}$ cm$^{-2}$ at the eastern edge of the outer western rim with a $T_{\rm k}\sim 40$ K and $n{\rm (H_2)}\sim 10^5$ cm$^{-3}$. Along the outer western rim, where the CO and the [C\,I] ($^3{\rm P}_2$-$^3{\rm P}_1$) line surface brightness maps peak, the column densities are $N{\rm (C\,I)}\sim 10^{16}$ cm$^{-2}$. The kinetic temperatures required to reproduce the intensities of the [C\,I] ($^3{\rm P}_1$-$^3{\rm P}_0$) and ($^3{\rm P}_2$-$^3{\rm P}_1$) lines through the outer rim are in the range $T_{\rm k}\sim 75-95$ K, hence, higher than those inferred from CO. These values are in good agreement with those inferred by \citet{Young97}.
}
\par{\citet{Young97} detected a high abundance of [C\,{\sc i}] relative to CO ($N$[C\,{\sc i}]$/N{\rm (CO)}\sim$ 6). We measured similar values at the edges of the outer western rim, but in the center of the rim, where the [C\,{\sc i}] and the CO surface brightness maps show the highest emission, the ratio can reach value up to $N$[C\,{\sc i}]$/N{\rm (CO)}\sim$ 9. The OH$^+$ emission peaks at the same position as [C\,{\sc i}], in the center of the outer western rim. The detection of OH$^+$ and [C\,{\sc i}] indicates an efficient OH$^+$ excitation mechanism and the ongoing photodissociation of the CO molecules by the central star's strong UV field or X-rays in the nebula's rims.}

\par{When CO is dissociated in an oxygen-rich planetary nebula with a C/O ratio close to 1, the free carbon atoms, if ionized (C$^+$) could form CH$^+$ (see \citet{Nagy13} for the Orion Bar case). However the CH$^+$ $J=1-0$ line at 835.17 GHz is not detected in the Helix, although it is present in bright carbon-rich PNs such as NGC\,7027, where CH$^+$ formation is dominated by reaction of vibrationally excited H$_2$ with C$^+$ \citep{Cernicharo97,Agundez10}. The reason for the absence of CH$^+$ might be a C$^+$ deficit, because other reactions destroy C$^+$ very efficiently. Other carbon-bearing molecules have been detected in the Helix Nebula, such as HCN, HCO$^+$, CN \citep{Bachiller97}, H$_2$CO, CCH and C$_3$H$_2$ \citep{Tenenbaum09}. These molecules survive despite the strong radiation field, shielded inside the clumps composed of gas-phase molecules mixed with dust, with densities as high as $10^5$ cm$^{-3}$ \citep{Howe94}. H$_2$CO, OH, and H$_2$O are easily formed in the dense photodissociation regions of protoplanetary nebulae \citep{Cernicharo89,Cernicharo04}. Our SPIRE FTS data do not cover the OH lines but indicate abundance for H$_2$O from its fundamental line at 557 GHz. The lack of emission from high excitation lines suggests that H$_2$O is restricted to the densest and cold regions, and that it could either be a remnant of the AGB phase of the star or to be produced by similar mechanisms to those proposed for CRL~618 by \citet{Cernicharo04}, which also lead to the formation of OH and H$_2$CO. 
}
\section{Summary and Conclusions}
\indent\par{We have reported the first detection of extended OH$^+$ line in emission in a planetary nebula. Independently and simultaneously, OH$^+$ lines in emission have been detected in several oxygen-rich planetary nebula, by \citet{Aleman13}, observed as part of the HerPlans project. Their work is also presented in this volume. 
}
\par{{\it Herschel} SPIRE FTS spectra display several atomic and molecular emission lines along the western rims of the Helix Nebula. The intensity maps of the atoms and molecules detected trace the dissociation of CO molecules and the stratification of the PDR along the outer western rims. CO arises from dense and shielded clumps in the western rims of the Helix Nebula. OH$^+$ and [C\,{\sc i}] likely trace the clumps surface where molecules reform after being photodissociated. Both, the OH$^+$ and [C\,{\sc i}] distributions are spatially coincident peaking at the same position in the outer western rim. [N\,{\sc ii}] traces the diffuse ionized gas in the interclump medium.}

\begin{acknowledgement}
We thank ASTROMADRID for funding support through the grant S2009ESP-1496, the consolider programme ASTROMOL: CSD2009-00038 and the Spanish MINECO (grants AYA2009-07304 and AYA2012-32032). FK is supported by the FWF project P23586 and the ffg ASAP project HIL. PvH and PR acknowledges support from the Belgian Science Policy Office (Belspo) through the ESA PRODEX program. HIPE is a joint development by the {\it Herschel} Science Ground Segment Consortium consisting of ESA, the NASA {\it Herschel} Science Center, and the HIFI, PACS, and SPIRE consortia. SPIRE has been developed by a consortium of institutes led by Cardiff University (UK) and including University of Lethbridge (Canada); NAOC (China); CEA, LAM (France); IFSI, Univ. Padua (Italy); IAC (Spain); Stockholm Observatory (Sweden); Imperial College London, RAL, UCL-MSSL, UKATC, Univ. Sussex (UK); and Caltech, JPL, NHSC, Univ. Colorado (USA). This development has been supported by national funding agencies: CSA (Canada); NAOC (China); CEA, CNES, CNRS (France); ASI (Italy); MCINN (Spain); SNSB (Sweden); STFC, UKSA (UK); and NASA (USA).
\end{acknowledgement}

\end{document}